# A 100 mA Low Voltage Linear Regulators for Systems on Chip Applications Using 0.18 µm CMOS Technology

Krit Salah-ddine[1], Zared Kamal[2], Qjidaa Hassan[3] and Zouak Mohcine[4]

[1]University Ibn Zohr Agadir Polydisciplinary Faculty of Ouarzazate Morocco.
*krit_salah@yahoo.fr*

[2] University Sidi Mohamed Ben Abdellah, Technical and Sciences Faculty,
Laboratory of Signals,
Systems and Components, Morocco.
*zaredk@hotmail.com*

[3] University Sidi Mohamed Ben Abdellah, Faculty of Sciences Dhar El Mehraz Fez
Laboratory of Electronic Signals –Systems and Informatics (LESSI), Morocco.
*qjidah@yahoo.fr*

[4] University Sidi Mohamed Ben Abdellah, Technical and Sciences Faculty,
laboratory of Signals, Systems and Components, Morocco.

**Abstract**
A novel design for a low dropout (LDO) voltage regulator is presented and dedicated to power many sections of a typical cellular handset. However, these baseband, RF, and audio sections have different requirements that influence which LDO is most appropriate. After discussion of the specific requirements, different LDOs are recommended. Also, some LDO design techniques are briefly discussed to demonstrate how an LDO may be optimized for a specific level of performance.
Cellular phone designs require linear regulators with low-dropout, low-noise, high PSRR, low quiescent current (Iq), and low-cost. They need to deliver a stable output and use small-value output capacitors. Ideally, one device would have all these characteristics and one low-dropout linear regulator (LDO) could be used anywhere in the phone without worry. But in practice, the various cell phone blocks are best powered by LDOs with different performance characteristics. This paper provides a new design methodology to choosing the right LDO to power each cell phone and especially for the Voltage Phase-Locked loops (VPLLs) blocks. Fabricated in a 0.18 µm CMOS process, the measured results show the adopted topology achieves a better phase noise than the conventional saturation current source. and the spread of the current limitation (without matching) is 100mA, the VPLLs system demonstrates a phase noise of 782 nv/sqrtHz at 100-kHz, and 33 nv/sqrtHz at 1 MHz, while quiescent current 33 µA from a 2.6 V supply voltage.

*Key words:*
*LDO, PSRR, low noise, cell phone, handset, RF, baseband, audio, GSM*

## 1. Introduction

Low dropout regulators (LDOs) are widely used and implemented in most circuit applications to provide regulated power supplies. The increasing demand of performance is especially apparent in mobile battery-operated products, such as cellular phones, pagers, camera recorders, and laptops [1-7]. For these products, very high PSRR, low noise regulators are needed. Moreover such high-performance regulators have to be designed in standard low-cost CMOS process, which makes them difficult to realize. For PSRR point of view, as depicted in [2], this kind of regulator requires a first-stage amplifier with a large gain-bandwidth product (Product of its dc-gain and cut-off frequency, which is typically 10 MHz). This first-stage amplifier performance can be achieved either by a large dc-gain, or by a high cut-off frequency.
Compared with switching regulators, LDOs are less expensive, smaller in size and easier to be used. Moreover, the noise of output voltage is lower and the response to input voltage transient and output load transient is faster. These advantages make LDOs suitable for battery-powered equipments, communication systems, portable systems, and post regulators of switching regulators. Among possible process technologies, CMOS technology is very attractive for LDO circuit implementation because of its low cost, low power consumption and potential for future system-on-chip integration.

In this paper, a CMOS LDO using new scheme that can maintain system stability with a maximum load current limitation =100 mA is proposed figure1 shows the regulator integrated in the system on chip and figure2 shows the schematic of the proposed regulator with power transistor who provide 100mA.

Moreover, it can provide stable output under all load conditions with a value of load capacitor equal 2.2 uF. The ESR of the load capacitor can range from zero to some finite value equal 100 mohom.

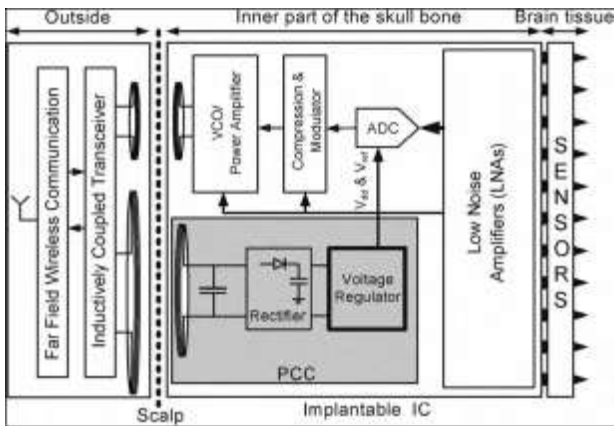
Fig.1 regulator integrated in the system on chip

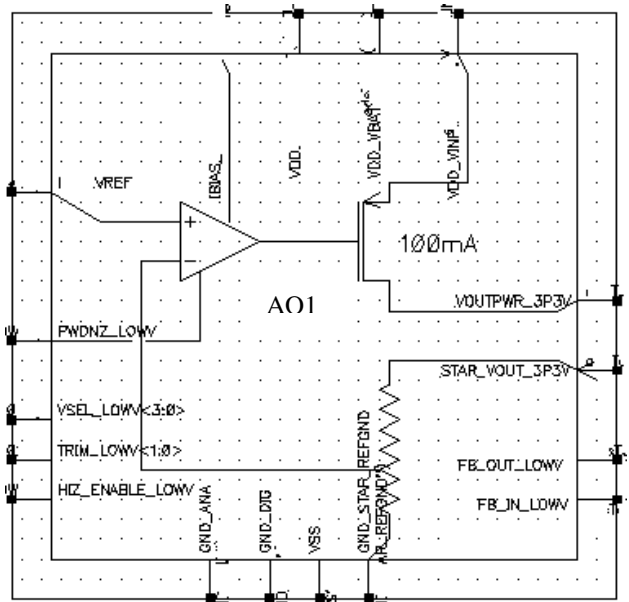
Fig.2 schematic of the proposed regulator with power transistor who provide 100mA.

## 2. Ldo Characterizations

LDO design involves three primary aspects, namely, regulating performance, current efficiency, and operating voltage [I]. These design aspects are explicitly stated in the following design specifications: I ) dropout, 2) line regulation, 3) load regulation, 4) temperature dependence, 5) transient output voltage variation as a result of load current steps, 6) quiescent current, and 7) power supply rejection ratio [I]. Table 1 lists the target specifications for the three LDO architectures. Each of these specifications is discussed in this section.

### i. Dropout
Dropout is the minimum input to output voltage difference at which the LDO ceases to regulate, determined by unacceptable decrease in output voltage. If the maximum current and minimum dropout conditions are not satisfied within the error margin, the LDO is then not performing its regulating function properly. The pass device must be large enough to guarantee the minimum dropout (source-to-drain or input-to-output voltage difference) while providing the maximum load current. Dropout is simulated by performing a dc sweep on the input voltage and plotting the output voltage. Dropout is the voltage difference between the input voltage and the output voltage at the point where the input voltage is minimum (2.6 V in this case). This point is estimated to be at the minimum input voltage at which the LDO is allowed to operate.

### ii. Line Regulation
Line regulation is the output voltage change as a result of a specific change in input voltage at a specific load current. Line regulation is simulated by performing a dc sweep on the input voltage and plotting the output voltage, is measured at both the maximum and minimum load currents.

### iii. Load Regulation
Load regulation is the ratio of the change in output voltage to the change in load current, which is the regulator output resistance, $R_{LDo}$. Load regulation is simulated by performing a dc sweep on the load current and plotting the output voltage. Then the load regulation of the regulator is $RoLm = 0,001 mV$

### iv. Temperature Dependence
Temperature dependence is the change in the output voltage due to a change in temperature. LDO temperature dependence is a function of the temperature dependence of the reference voltage and the offset voltage of the error amplifier. The temperature coefficient of a CMOS band-gap voltage reference can be as low as 15 ppm/°C over a temperature range from -40°C to 125°C [2]. Therefore, the temperature dependence of the band-gap reference voltage is ignored in this simulation. LDO temperature dependence is simulated with a dc sweep on the temperature and plotting the output voltage.

### v. Transient Output Voltage Variation

Transient output voltage variation is the output voltage change in response to transient load current variation and is a function of four parameters: system time response, output capacitance, maximum load current, and ESR of the output capacitor. Transient output voltage variation is simulated by applying a transient load current signal and plotting the transient output voltage, is measured for a certain At (rising and falling time of the transient load current signal).

### vi. Quiescent Current

Quiescent current is the total current drawn from the voltage supply at zero load current. Quiescent current is simulated by performing a dc operating point and measuring the total current consumed by the EA and the feedback resistors.

### vii. Power Supply Rejection Ratio

Power supply rejection ratio (PSRR) is the ratio of the change in output voltage to the change in input voltage supply. It is also defined as the ac voltage gain from the input node to the output node of the LDO regulator. PSRR is simulated by performing an ac sweep of the input voltage supply and plotting the ratio of the output voltage to the input voltage. PSRR is measured at the frequencies of interest in dB

## 2. Design of Current Limiting Circuit

2.1 Design Requirement of Current Limiting Circuit

A current limiting circuit used in LDO linear voltage regulator should at least meet the requirements as follows:
a. When overcurrent hasn't taken place, the voltage regulator should regulate the output voltage Vout normally, and the current limiting circuit should have little effect on it.
b. A current limiting circuit should first include output current detecting devices or block to detect if output current IO has exceeded the maximum rated value.
c. After the current limiting circuit starts up, it should cut off the negative feedback loop of the regulator. Then the regulator cannot regulate the output voltage any more.
d. After foldback current limiting circuit starts up, IO (Input Output) will decrease as Vout decreases. As the output is shorted, IO will be limited to a value much less than the maximum rated value.
Besides, a good current limiting circuit should take some other factors into consideration, such as: low quiescent current and power consumption, few devices, low cost, and soon.

2.2 Design Principle of Current Limiting Circuit

The current limiting circuit presented in the paper is showed in Fig. 1. It comprises output current sampling circuit, constant current limiting circuit and foldback current limiting circuit. Signals VB1 and VB2 are generated by the self-biasing circuit of the error amplifier (We only give the second stage of the amplifier in Fig. 1). The potential of VB2 is constant, and VSG_MP＝VDD－VB1 holds constant as well.
MN1 and MP2 make up of the second stage of the error amplifier. AO1 is its input as well as the output of the first stage and AMP_OUT is its output as well as the output of the error amplifier. PW is pass element. PWDNZ is the enable control signal. When it is at high potential, MP1 is off and the circuit works normally.

## 3. Simulation and experimental Results of the proposed Voltage Regulator

the architecture of the proposed voltage regulator as shows in Figure 3 which reduces the total cost and facilitates the regulator implantation. The supply voltage denoted as Vin is provided by the rectifier output, and can be as low as 2.6V. the reference voltages VREF =0,75V as bandgap reference circuit [5] with dynamic start-up and turn-on time circuitry is used to generate the required reference voltages and currents the simulation results is shown in figure 4. The bandgap is supplied from the regulator output, which mitigates the need for high PSRR reference voltage generation.
the current limitation of the proposed regulator LDO was simulated and realized from the structure depicted in fig.3, using a 2.2μF external capacitor on $V_{OUT}$. This regulator was designed to deliver 3.40V with a maximum load current of 100mA.
Figure 5 shows the simulation results of the output noise and PSRR outputs, when the $V_{DD}$ voltage is rising and falling. This indicates, as explained previously, that when $V_{DD}$ rises and is below 2.35V, the $P_{OR}$ signal stays low, and forces $V_{OUT}$ to follow $V_{DD}$. In these conditions, the total quiescent current for this circuit is below 1μA. For higher values of $V_{DD}$, the LDO regulates the output voltage $V_{OUT}$ to 2.40V. The maximum quiescent current is obtained for $V_{DD}$=5.5V, and is equal to 1.5μA.
The figure 6 shows that the the phase margin versus iload who verify the stability of our architecture. Figure 7 schows the dc-ligne regulation of the regulator.
Then this ultra-low quiescent regulator including the POR was fabricated in a CMOS 0.18 μm process. It has been optimized for quiescent current.
The external load capacitor was 2.2μF, the output load current is less or equal than 1mA, and the input voltage $V_{DD}$ is below 5.5V.

The measured total quiescent current was less than 2μA for $V_{DD}$ in the 0V-5.5V range. These measured values are in good accordance with the simulated results above.

Table 1: Margin specifications

| *Parameters* | *Typic Value* | Units |
|---|---|---|
| Vin | 2.6 | V |
| Ilimit | 100 | mA |
| Ligne R | 5 | mV |
| Load R | 20 | mV |
| PSRR | 65 | dB |
| DC gain | 89 | dB |

Table 2: Performance comparison between recent works on SoC LDOs

|  | [14] | [15] | [16] | [17] | This work |
|---|---|---|---|---|---|
| Year | 2003 | 2007 | 2007 | 2009 | 2011 |
| Process | 0.6μ | 0.35μ | 0.35μ | 0.35μ | 0.18μ |
| Vin | 1.5V | 3V | 1.2-3.3 V | 1.2-1.5 V | 2.6 V |
| Vout | 1.3 V | 2.8 V | 1V | 1V | 2.4 V |
| $I_Q$ | 38 μA | 65 μA | 100 μA | 45 μA | 1.5 μA |
| $I_L^{MAX}$ | 100 mA | 50 mA | 100mA | 50mA | 100mA |
| ΔVout | 100mV | < 90mV | 50mV | 70mV | 20mV |

### 4. Performance Comparison

Table 2 provides comparison between the performance of the proposed LDO regulator and other published designs that are targeted for SoC power management.

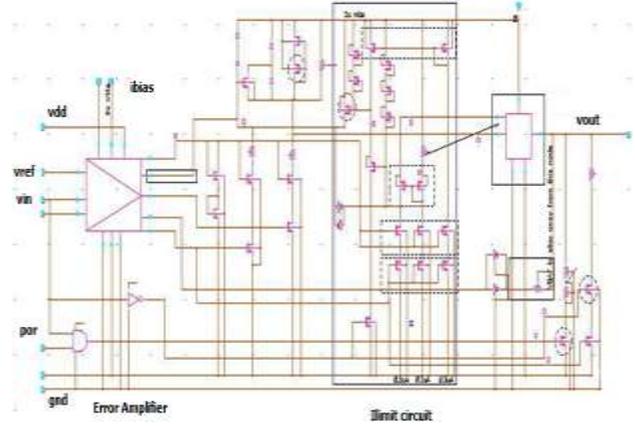

Fig.3 Architecture of the proposed regulator

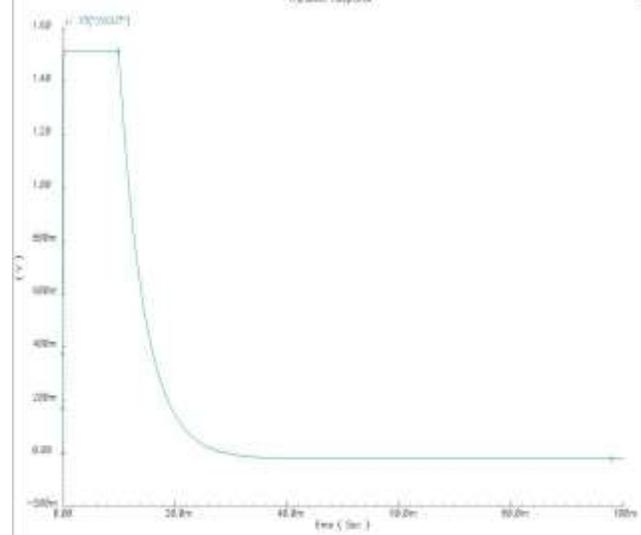

Fig.4 Startup and Turn-Off

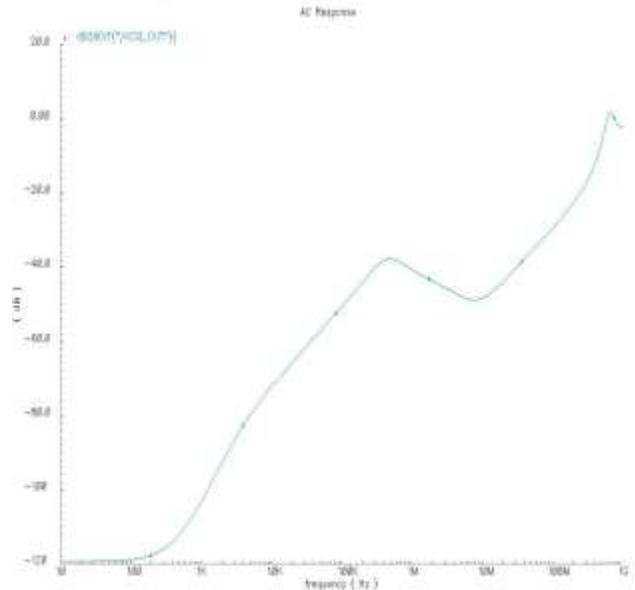

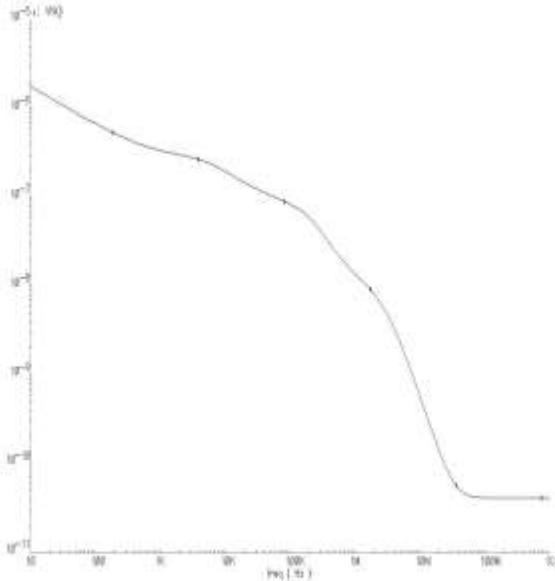

Fig.5 Output Noise and PSRR

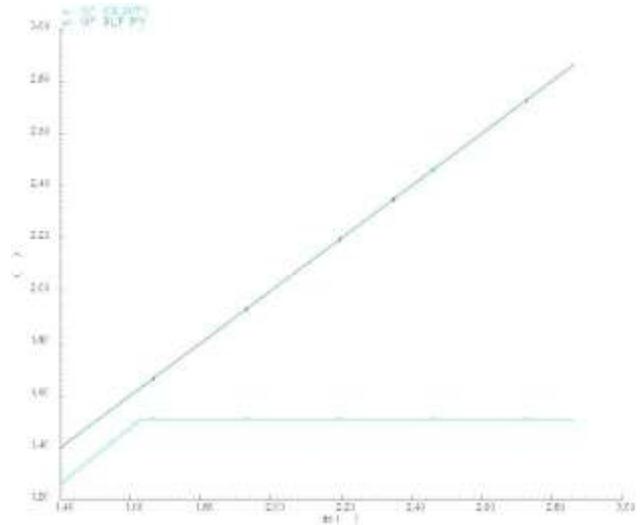

Fig.7 Dc ligne Regulation

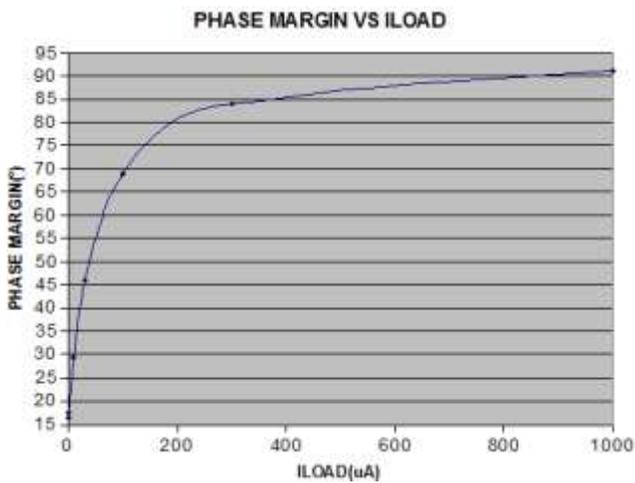

Fig.6 Phase margin versus iload

## 5. Abreviations

| abbreviations | Sens |
|---|---|
| PSRR | Power Supply Rejection Ratio |
| VPLLs | Voltage Phase-Locked Loops |
| CMOS | Complementary Metal-Oxyde Semiconductor |
| LDO | Low Drop Out |
| ESR | Equivalent Series Resistor |
| DC | Direct Current |
| Iload | Current Load |
| VB | FeedBack |
| VDD | Voltage positive supply |
| VREF | Voltage Reference |
| VOUT | Output Voltage |
| PWDNZ | Power Down Zero: this input can be driven low (at a logic 0) |
| POR | Power On Reset |
| SoC | System on Chip |
| $I_Q$ | Quiescent Current |

Table 3: Abbreviations of the words used in this proposed work.

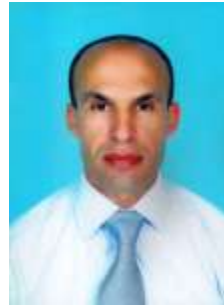

**Salah-ddine Krit** received the B.S. and Ph.D degrees in Microectronics Engineering from Fes Sidi Mohammed Ben Abdellah university, Fez, Morroco. Institute in 2004 and 2009, respectively. During 2002-2008, he is also an engineer Team lead in audio and power management Integrated Circuits (ICs) Research.

Design, simulation and layout of analog and digital blocks dedicated for wireless sensor networks (WSN) and satellite communication systems using CMOS technology. He is currently a professor with Polydisciplinary Faculty of Ouarzazate, Ibn Zohr university, Agadir, Morroco. His research interests include wireless sensor Networks (Software and Hardware), microtechnology and nanotechnology for wireless communications.

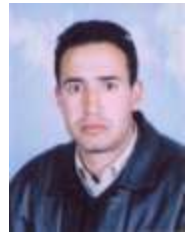

**Zared kamal** received the B.S. and M.S. degrees in Electrical Engineering from faculty of science Dhar El Mehraz Fez in 1997 and 2004, respectively. During 2008-2010 PhD researchers in electrical engineering, 2004-2010, He is currently teacher of sciences computer.

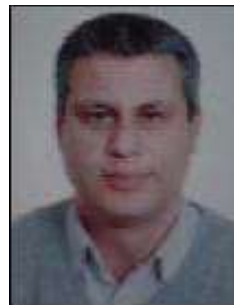

**Hassan Qjidaa** received his M.Sc.and PhD in Applied Physics from Claude Bernard University of Lyon France in 1983 and 1987 respectively. He got the Pr. degree in Electrical Engineering from Sidi Mohammed Ben Abdellah university, Fès , Morroco 1999.

He is now an Professor in the Dept. of Physics in Sidi research interests include Very-large-scale integration (VLSI) solution, Image manuscripts Recognition, Cognitive Science, Image Processing, Computer Graphics, Pattern Recognition, Neural Networks, Human-machine Interface, Artificial Intelligence, Robotics and so on.

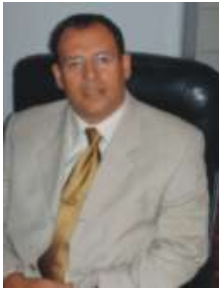 **Mohcine Zouak** was born in Morocco on 1963. He received the "Docteur d'Etat" degree in radar signal processing from Sidi Mohamed Ben Abdellah University, Fez (Morocco) in 1995 and Ph.D degree in electronics and informatics systems from the University of Nantes (France).

He is currently a professor with Science and Technical Faculty, Fez (Morocco), where he manages the UFR of Signals, Systems and Components. His research interests include sensors array processing, signal processing for wireless communications, and statistical signal processing. Since 2005, he has also the dean of Science and Technical Faculty, Fez.